\documentclass[aps,prb,twocolumn,groupedaddress,showpacs,superscriptaddress]{revtex4-1}
\usepackage{graphicx}
\usepackage{array}
\usepackage{multirow}
\usepackage{amsmath}
\usepackage{amssymb}
\usepackage{gensymb}
\usepackage{graphics}
\usepackage{color}
\usepackage{multirow}
\begin{document}

%\begin{CJK*}{GBK}{}

%\date{\today}
\title{Probing IrTe$_2$ crystal symmetry by polarized Raman scattering}

\author{N. Lazarevi\'c}
 \affiliation{Center for Solid State Physics and New Materials, Institute of Physics Belgrade, University of Belgrade, Pregrevica 118, 11080 Belgrade, Serbia}

\author{E. S. Bozin}
\affiliation{Condensed Matter Physics and Materials Science Department, Brookhaven
National Laboratory, Upton, New York 11973-5000, USA}

\author{M. \v{S}\'cepanovi\'c}
 \affiliation{Center for Solid State Physics and New Materials, Institute of Physics Belgrade, University of Belgrade, Pregrevica 118, 11080 Belgrade, Serbia}

\author{M. Opa\v{c}i\'c}
 \affiliation{Center for Solid State Physics and New Materials, Institute of Physics Belgrade, University of Belgrade, Pregrevica 118, 11080 Belgrade, Serbia}

\author{Hechang Lei}
\affiliation{Condensed Matter Physics and Materials Science Department, Brookhaven
National Laboratory, Upton, New York 11973-5000, USA}

\author{C. Petrovic}
\affiliation{Condensed Matter Physics and Materials Science Department, Brookhaven
National Laboratory, Upton, New York 11973-5000, USA}

\author{Z. V. Popovi\'c}
\affiliation{Center for Solid State Physics and New Materials, Institute of Physics Belgrade, University of Belgrade,
Pregrevica 118, 11080 Belgrade, Serbia}

\date{\today}

\begin{abstract}
Polarized Raman scattering measurements on IrTe$_2$ single crystals carried out over 15~K - 640~K temperature range, and across the structural phase transition, reveal new insights regarding the crystal symmetry.\ In the high temperature regime three Raman active modes are observed at all studied temperatures above the structural phase transition, rather than two as predicted by the factor group analysis for the assumed $P\bar{3}m1$ symmetry. This indicates that the actual symmetry of the high temperature phase is lower than previously thought.\ Observation of an additional E$_g$ mode at high temperature can be explained by doubling of the original trigonal unit cell along the $c$-axis and within the $P\bar{3}c1$ symmetry.\ In the low temperature regime (below 245~K) the new Raman modes appear as a consequence of the symmetry lowering phase transition and corresponding increase of the primitive cell.\ All the modes observed below the phase transition temperature can be assigned within the monoclinic crystal symmetry.\ Temperature dependence of the Raman active phonons in both  phases are mainly driven by anharmonicity effects. The results call for reconsideration of the crystallographic phases of IrTe$_2$.
\end{abstract}

\maketitle
%\end{CJK*}

\section{Introduction}

Although known for some time,\cite{Jobic1991,Jobic1992} the interest in IrTe$_2$ has been renewed recently with the discovery of superconductivity.\cite{Yang,Pyon,Ootsuki,Kamitani} By doping this layered compound with Pt, Pd and Cu, the phase transition which occurs at low temperatures\cite{Trig-Mono} is suppressed and superconductivity emerges.\cite{Yang,Pyon,Ootsuki,Kamitani,kiswa;prb13}
At room temperature IrTe$_2$ has a trigonal symmetry with edge-sharing IrTe$_6$ octahedra forming layers stacked along the $c$-axis,\cite{Trig-Mono} as shown in Figure~\ref{IrTe2-struc}.\ As temperature is decreased the system undergoes a symmetry lowering phase transition in temperature range between 220 K and 280 K, with the exact transition temperature, $T_{PT}$, presumably depending on the sample form (powder vs single crystal) and the thermal cycle details (cooling or warming).\cite{Trig-Mono,Yang,Pyon,Ootsuki,Kamitani,Depolymerization} The phase transition is accompanied by a hump in electrical resistivity and a drop in magnetic susceptibility,\cite{Fang} anomalies reminiscent to these associated with an onset of the CDW state observed in other TX$_2$ systems.\cite{Cao} However, the exact nature of the low temperature phase remains controversial, since no signatures of the CDW gap in IrTe$_2$ have been seen in angle resolved photoemission and optical spectroscopy studies.\cite{Fang,Ootsuki,Depolymerization} Recent band structure calculations combined with x-ray absorption spectroscopy measurements suggest that the dramatic change in the interlayer and intralayer hybridizations could play an important role in the structural phase transition of IrTe$_2$.\cite{Kamitani} More recently, it has also been suggested that the depolymerization of the polymeric Te-Te bonds might be responsible for the structural phase transition.\cite{Depolymerization}

Although prior crystallographic analyses showed that the IrTe$_2$ crystal structure changes from trigonal to monoclinic with decreasing temperature, the low temperature structure is still subject of debate.\cite{Trig-Mono} It was argued that the initially assigned monoclinic $C2/m$ symmetry cannot fully describe the structure below the phase transition.\cite{Zhang,P1,P-1} Consequently the proposed crystal symmetry was further lowered down to triclinic $P\bar{1}$\cite{P1} and even $P1$\cite{P-1}.\ Moreover, it was also suggested that the trigonal and monoclinic structures coexist intrinsically below the phase transition.\cite{Zhang} The nature of the phase transition as well as the symmetry of the low temperature phase therefore still remain open questions.

Important information concerning the symmetry of the crystal system can be obtained by utilizing the properties of Raman spectroscopy and by performing the measurements in different polarization configurations whereby one can probe different scattering channels.\ Raman spectroscopy also emerges as a valuable tool for detecting the intrinsic phase separation.\cite{LazarevicVacancy}

Here we present results of a systematic Raman scattering study on IrTe$_2$ single crystals.\ The spectra were collected in different scattering geometries at various temperatures.\ The room temperature Raman spectra were analyzed within the trigonal crystal symmetry.\ Three instead of two peaks, which are predicted by the factor group analysis (FGA) for the $P\bar{3}m1$ space group, are observed in the Raman spectra suggesting different crystal symmetry of IrTe$_2$ in the high temperature phase from that previously assumed.\ The same phonon structure persists at $T \gg T_{PT}$, indicating that it is a true characteristics of the high temperature phase.\ At temperature below $T_{PT}=$ 245 K the clear fingerprint of the first order structural phase transition is observed in the Raman spectra.\ The observed modes are interpreted within the monoclinic crystal symmetry.\ No signatures of the trigonal unit cell presence have been detected in the low temperature Raman scattering spectra.\ All temperature induced effects in both phases are mostly anharmonic. These observations provide important insights and constraints for possible crystal symmetries of this system in different temperature regimes.

\section{Experiment}

Single crystals of IrTe$_2$ were prepared by the  self-flux method.\ Ir and Te were mixed in 18:82 stoichiometric ratio, heated in alumina crucibles under and Ar atmosphere up to 1160\celsius, kept at that temperature for 24 hours and than cooled to 400\celsius~ over 130 hours.\ Excess Te flux was removed at 400\celsius~ by centrifugation.\ Plate-like mm-size crystals were obtained.\
Magnetization and resistivity data were measured by warming the sample from 5 K.\ They are in good agreement with published values.\cite{Yang}

Raman scattering measurements were performed using a JY T64000 Raman system with the 1800/1800/1800 groves/mm gratings and TriVista 557 Raman system with the 900/900/1800 groves/mm gratings combination, both in backscattering micro-Raman configuration.\ The 514.5 nm line of a mixed Ar$^+$/Kr$^+$ gas laser was used as an excitation source.\ High temperature measurement were preformed in Ar environment by using Linkam THGS600 heating stage.\ Low temperature measurements were performed using KONTI CryoVac continuous flow cryostat with 0.5 mm thick window in warming regime.\ Complementary atomic pair distribution function (PDF) measurement at 300~K was performed on finely pulverized sample at X17A beamline of National Synchrotron Light Source (NSLS) at Brookhaven National Laboratory, utilizing 67.42~keV X-ray beam within commonly used rapid-acquisition setup~\cite{chupa;jac03} featuring sample to detector distance of 204.25~mm, and with access to a wide momentum transfer range up to 28~\AA$^{-1}$. Standard corrections, PDF-data processing and structure modeling protocols were utilized, as described in detail elsewhere.~\cite{egami;b;utbp12}.

\section{Results and discussion}

%%%%%%%%%%%%%%%%%%%%%%%%%%%%%%%%%%%%%%%%%%%%%%%%%%%%%%%%%%%%%%%%%%%%%%%%%%%%
\begin{table*}[htb!]
\caption{Considerations of Raman tensors for three different crystal systems and the Raman mode distribution in the $\Gamma$ point and various space groups of interest for IrTe$_2$.}
\label{tab.1}
\begin{ruledtabular}
\centering
\begin{tabular}{cccc}
   Crystal system & \multicolumn{2}{c}{Raman tensors\cite{kuzmany}} & Raman modes\\ \hline \noalign{\smallskip}
  %%%%%%%%%%%%%%%%%%%%%%%%%%%%%%%%%%%%%%%%%%%%%%%%%%%%%%%%%%%%%%%%%%%%%%%%%%%%%%%%%%%%%%%%%%%%%%%%%%%%%%%%%%%%%%%%%%%%%%%%%%%%%%%%%%%
      Trigonal ($O_z\parallel C_3$ $O_y\parallel C_2$) &    $\hat{R}_{A_{1g}}$=$\left(\begin{matrix}
                                    a & 0 & 0 \\
                                    0 & a & 0 \\
                                    0 & 0 & b%
                                    \end{matrix} \right)$                    &  $\hat{R}_{E_{g}}$=$\left(\begin{matrix}
                                                                                -c & 0 & 0 \\
                                                                                0 & c & 0 \\
                                                                                0 & 0 & 0%
                                                                                \end{matrix} \right)$, $\left(\begin{matrix}
                                                                                0 & c & 0 \\
                                                                                c & 0 & 0 \\
                                                                                0 & 0 & 0%
                                                                                \end{matrix} \right)$
                                                                                            &\begin{tabular}{@{}c@{}}
                                                                                             $\Gamma_{P\bar{3}m1}$=A$_{1g}$+E$_{g}$ \\
                                                                                             $\Gamma_{P\bar{3}c1}$=A$_{1g}$+A$_{2g}(silent)$+2E$_{g}$ \\
                                                                                             \end{tabular}\\ \noalign{\smallskip}  \noalign{\smallskip}
  %%%%%%%%%%%%%%%%%%%%%%%%%%%%%%%%%%%%%%%%%%%%%%%%%%%%%%%%%%%%%%%%%%%%%%%%%%%%%%%%%%%%%%%%%%%%%%%%%%%%%%%%%%%%%%%%%%%%%%%%%%%%%%%%%%%
   Monoclinic ($O_y\parallel C_2$) &    $\hat{R}_{A_{g}}$=$\left(\begin{matrix}
                                    b & 0 & d \\
                                    0 & c & 0 \\
                                    d & 0 & a%
                                    \end{matrix} \right)$                    &  $\hat{R}_{B_{g}}$=$\left(\begin{matrix}
                                                                                0 & f & 0 \\
                                                                                f & 0 & e \\
                                                                                0 & e & 0%
                                                                                \end{matrix} \right)$
                                                                                            &\begin{tabular}{@{}c@{}} $\Gamma_{C2/m}$=2A$_g$+B$_g$

                                                                                            \end{tabular}\\
                                                                                \noalign{\smallskip}  \noalign{\smallskip}
   %%%%%%%%%%%%%%%%%%%%%%%%%%%%%%%%%%%%%%%%%%%%%%%%%%%%%%%%%%%%%%%%%%%%%%%%%%%%%%%%%%%%%%%%%%%%%%%%%%%%%%%%%%%%%%%%%%%%%%%%%%%%%%%%%%%
  Triclinic & \multicolumn{2}{c}{
                                    $\hat{R}_{A}$=$\hat{R}_{A_{g}}$=$\left(\begin{matrix}
                                    a & d & e \\
                                    d & b & f \\
                                    e & f & c%
                                    \end{matrix} \right)$}
                                                                   & \begin{tabular}{@{}c@{}}
                                                                   $\Gamma_{P\bar{1}}$=21A$_g$\\
                                                                   $\Gamma_{P1}$=222A
                                                                   \end{tabular}\\
                                                                   \noalign{\smallskip}  \noalign{\smallskip}
  %%%%%%%%%%%%%%%%%%%%%%%%%%%%%%%%%%%%%%%%%%%%%%%%%%%%%%%%%%%%%%%%%%%%%%%%%%%%%%%%%%%%%%%%%%%%%%%%%%%%%%%%%%%%%%%%%%%%%%%%%%%%%%%%%%%
\end{tabular}
\end{ruledtabular}
\end{table*}
%%%%%%%%%%%%%%%%%%%%%%%%%%%%%%%%%%%%%%%%%%%%%%%%%%%%%%%%%%%%%%%%%%%%%%%%%%%%

%%%%%%%%%%%%%%%%%%%%%%%%%%%%%%%%%%%%%%%%%%%%%%%%%%%%%%%%%%%%%%%%%%%%%%%%%%%%
\begin{table}[htb!]
\caption{Structural parameters for the $P\bar{3}c1$ phase obtained from PDF analysis at 300~K.\ Lattice parameter are a=b=3.929(4) \AA,~ c=10.805(2) \AA. $U_{ij}$ (\AA$^2\times~10^3$) are nonzero components of the displacement tensor.}
\label{tab.2}
\begin{ruledtabular}
\centering
\begin{tabular}{cccccc}

         Atom & $x$ & $y$ & $z$ & $U_{11}=U_{22}$  & $U_{33}$ \\  \hline
         Ir   &   0 & 0   &  0 & 6(2)& 6(3) \\
         Te   & 1/3 & 2/3 & 0.126(1)& 7(2)& 9(4)\\
\multicolumn{6}{c}{$\chi^2$ = 0.002} \\
\multicolumn{6}{c}{ R$_{wp}$ =0.106}

\end{tabular}
\end{ruledtabular}
\end{table}
%%%%%%%%%%%%%%%%%%%%%%%%%%%%%%%%%%%%%%%%%%%%%%%%%%%%%%%%%%%%%%%%%%%%%%%%%%%%

\begin{figure}[t!]
\includegraphics[width=0.45\textwidth]{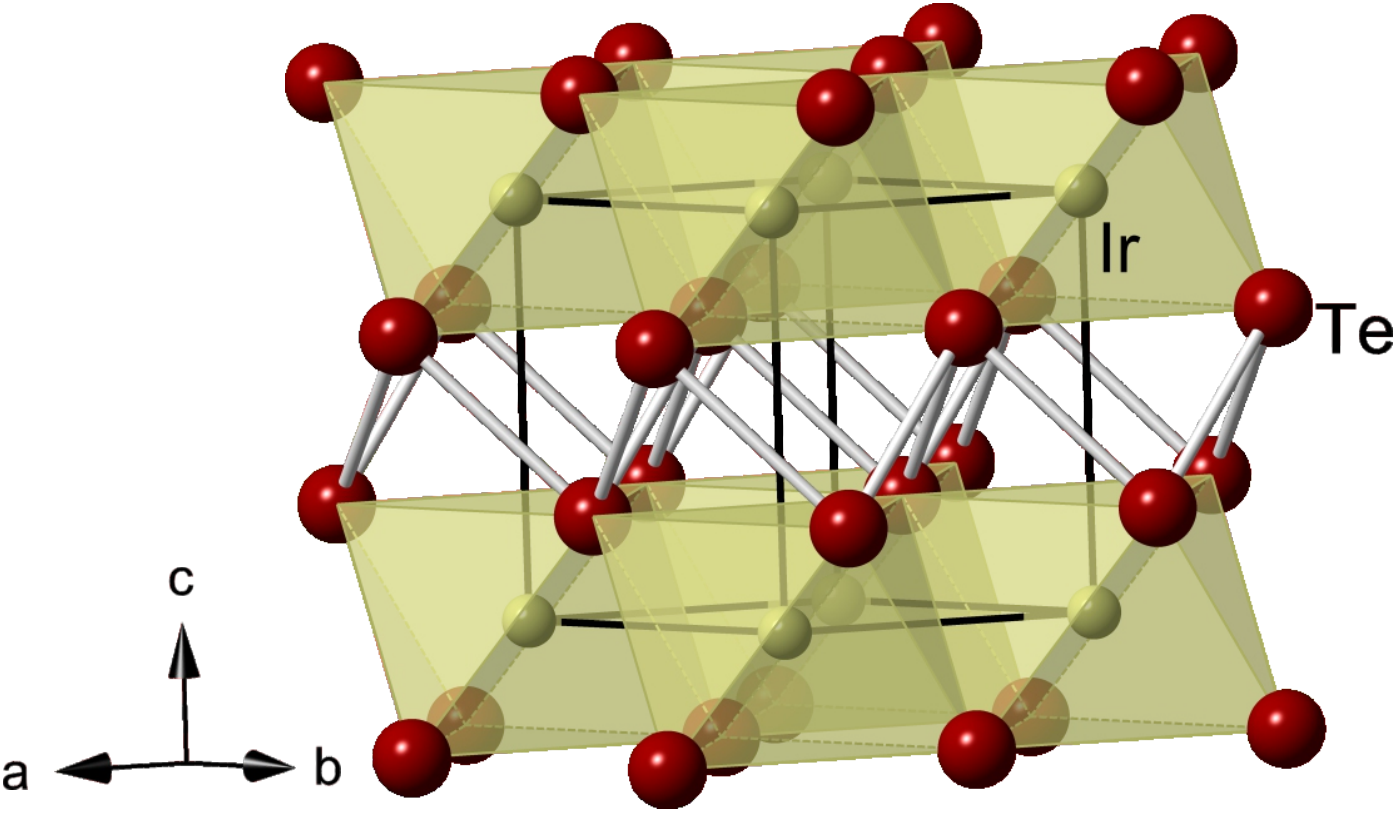}
\caption{(Color online) Crystal structure of IrTe$_2$ in the trigonal phase.\ Solid lines represent single $P\bar{3}m1$ unit cell, with yellow (red) spheres indicating positions of Ir (Te).}
\label{IrTe2-struc}
\end{figure}

IrTe$_2$ crystalizes in trigonal type of structure ($P\bar{3}m1$ space group) with one molecular unit per unit cell (Figure~\ref{IrTe2-struc}).\cite{Trig-Mono,Soulard2005} The crystal structure consists of IrTe$_2$ layers which are made up of edge sharing IrTe$_6$ octahedra.\ Short Te-Te bonds between adjacent IrTe$_2$ result in three-dimensional polymeric networks thereby reducing the $c/a$ ratio in comparison with the standard hexagonal closed packing of the CdI$_2$ structure.\cite{Jobic1991,Jobic1991,Canadell1992,Depolymerization} This is related to the Ir$^{+3}$ state and the fractional oxidation state of Te anions (Te$^{-1.5}$).\cite{Jobic1992,Depolymerization}

%
%%%%%%%%%%%%%%%%%%%%%%%%%%%%%%%%%%%
\subsection{High temperature phase}
%%%%%%%%%%%%%%%%%%%%%%%%%%%%%%%%%%%
%

\begin{figure}[t!]
\includegraphics[width=0.45\textwidth]{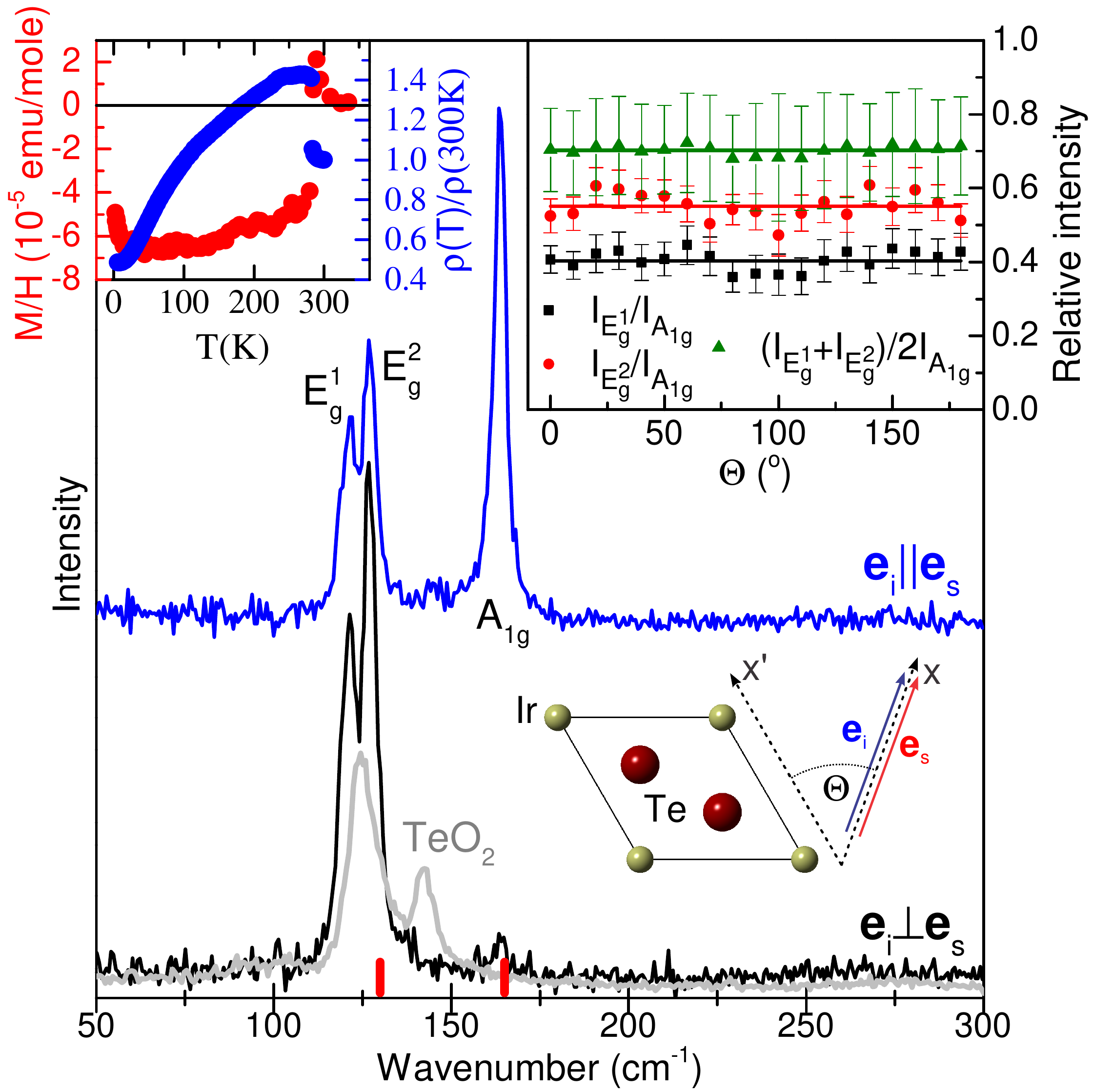}
\caption{(Color online)  Room temperature Raman scattering spectra of IrTe$_2$ measured using  JY T64000 Raman system in different polarization configurations.\ Grey lines represent the spectra of TeO$_2$ with scaled intensity.\ Red markers represent phonon energies at $\Gamma$ point calculated by Cao \emph{et al}.\cite{Cao} Inset on the left: Magnetization and resistivity data measured by warming the sample from the base temperature.\ Inset on the right: Relative intensities of the Raman active modes measured in parallel polarization configuration for different orientations of the sample with respect to the laboratory axis.}
\label{fig1}
\end{figure}

Figure~\ref{fig1} shows the room temperature polarized Raman scattering spectra of IrTe$_2$ single crystals measured from the (001) plane of the sample.\ Although the FGA for the $P\bar{3}m1$ space group predicts only two Raman active modes (A$_{1g}$+E$_{g}$) to be observed in the scattering experiment, three peaks are clearly distinguished in the data.\ Contribution to the Raman spectra originating from scattering on possible TeO$_2$ impurities can be safely excluded (see Figure~\ref{fig1}).\ According to the selection rules for the trigonal system, summarized in Table~\ref{tab.1}, the A$_{1g}$ mode can only be observed in parallel but not in crossed polarization configuration, whereas E$_g$ mode can be observed in both parallel and crossed polarization configurations.\ Consequently, the peak at about 164 cm$^{-1}$, which is indeed observed in parallel but not in cross polarization configuration, is attributed to the A$_{1g}$ symmetry mode.\ The energy of this mode is in very good agreement with the calculated value\cite{Cao} (red mark in Fig.~\ref{fig1}).\ Whereas the numerical calculations\cite{Cao} further predict a single E$_g$ mode at about 130 cm$^{-1}$ to be observed in the Raman scattering experiment, two peaks at about 121 cm$^{-1}$ and 126 cm$^{-1}$ are unambiguously observed in the data in this energy range (see Fig.~\ref{fig1}).\ These modes are observed in both parallel and cross polarization configurations, which suggests their E$_g$ symmetry.\ The appearance of two modes in the energy range where only one mode is expected indicates that the original crystal symmetry assignment for the high temperature phase may be inadequate, and that the actual symmetry is in fact lower.\ We consider the issue of the symmetry of high temperature phase in more detail next.

\begin{figure}[t!]
\includegraphics[width=0.45\textwidth]{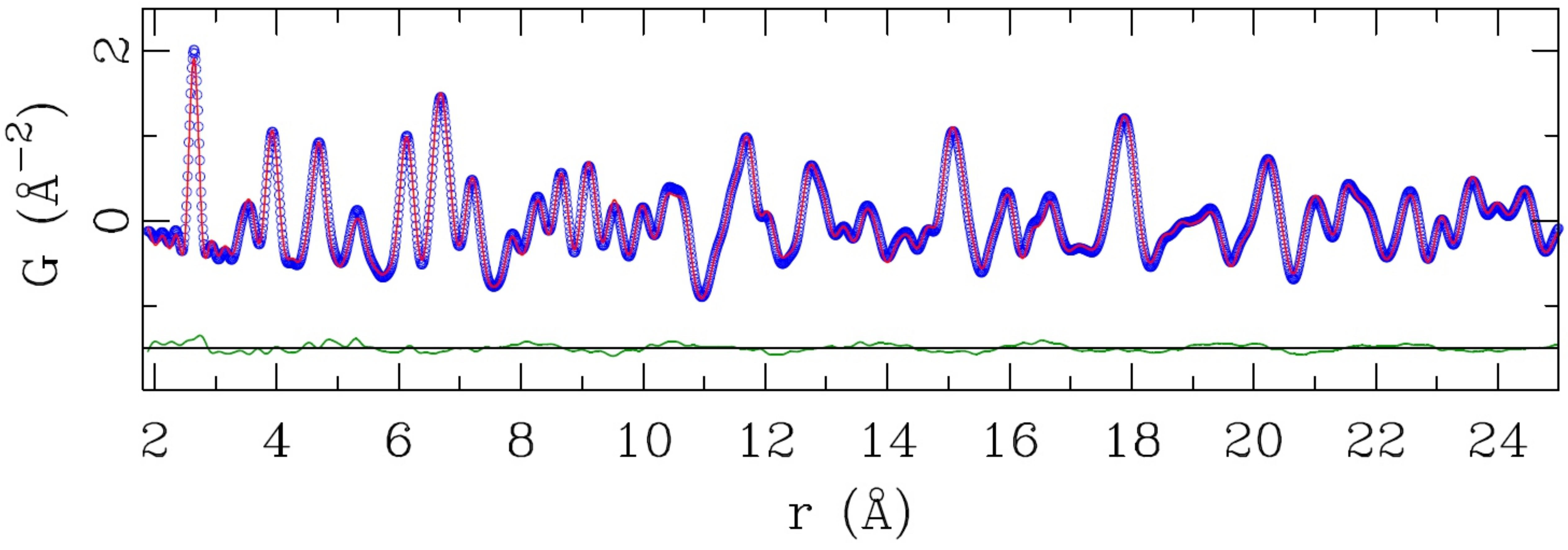}
\caption{(Color online)  Room temperature x-ray PDF of IrTe$_{2}$: experimental data (blue open symbols), $P\bar{3}c1$ model (red solid line),
and difference curve (green solid line) which is offset for clarity. Structural parameters are summarized in Table~\ref{tab.2}.}
\label{figPDF}
\end{figure}

The first possibility is that the IrTe$_2$ symmetry is lowered at room temperature to some t-subgroup of the $P\bar{3}m1$.\cite{international}\ This would imply splitting of a double degenerate E$_g$ mode into a A$_g$-B$_g$ doublet.\cite{fateley} Hereby obtained modes would display different angular Raman intensity dependencies as the sample orientation is varied in parallel polarization configuration (see Fig.~\ref{fig1}).\ On the contrary (as can be seen in the inset to Fig.~\ref{fig1}) both modes at 121 cm$^{-1}$ and 126 cm$^{-1}$ exhibit the same angular intensity dependence thereby excluding the possibility of the E$_g$ mode splitting \emph{i.e.} symmetry lowering to some t-subgroup of the $P\bar{3}m1$.\ Furthermore, it confirms the E$_g$ nature of the 121 cm$^{-1}$ and 126 cm$^{-1}$ modes, since, for a trigonal system (see Table~\ref{tab.1}), both A$_{1g}$ and E$_g$ mode intensities are independent on the sample orientation when measured in parallel polarization configuration.\

The second possibility which could explain the observed appearance of the two E$_g$ modes instead of a single E$_g$ mode is the symmetry change to some k-subgroup of $P\bar{3}m1$.\cite{international}\ The simplest option is $P\bar{3}c1$ (Z=2) with Ir atoms located on $2b$ site and Te atoms at $4d$ site.\ The $P\bar{3}c1$ unit cell is built by doubling of the $P\bar{3}m1$ unit cell along the $c$-axis (see Fig~\ref{IrTe2-struc}).\ The FGA for the $P\bar{3}c1$ predicts three Raman active modes to be observed in the Raman scattering experiment (A$_{1g}$+2E$_{g}$), which is in complete agreement with our findings.\ To further verify the plausibility of this assumption, we performed structural analysis of room temperature X-ray PDF data of IrTe$_2$ using $P\bar{3}c1$ model.\ Fit results are shown in Figure.~\ref{figPDF} and summarized in Table~\ref{tab.2}.

\begin{figure}
\includegraphics[width=0.45\textwidth]{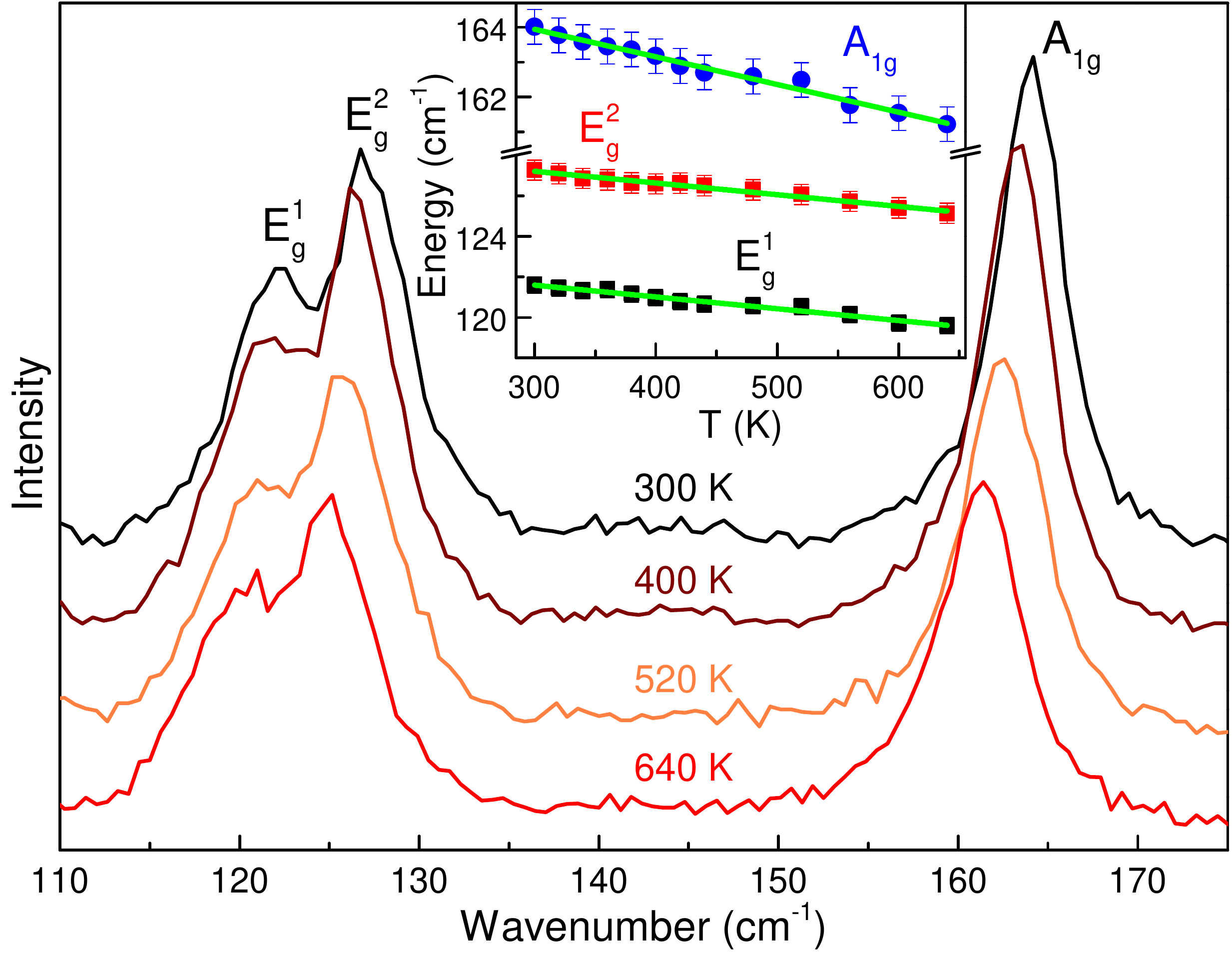}
\caption{(Color online) Raman scattering spectra of IrTe$_{2}$ single crystals measured at various temperatures, as indicated, in the high temperature regime using TriVista 557 Raman system.\ Inset: Energy temperature dependence of the A$_{1g}$, E$^{1}_{g}$ and E$^{2}_{g}$ Raman active modes.\ The solid green lines represent calculated spectra by using the standard three-phonon anharmonicity model,\cite{anh1} where ($\omega_0=$ 123.4(2) cm$^{-1}$, $C=$ 0.26(2) cm$^{-1}$), ($\omega_0=$ 129.0(2) cm$^{-1}$, $C=$ 0.27(2) cm$^{-1}$) and ($\omega_0=$ 160.4(2) cm$^{-1}$, $C=$ 0.48(3) cm$^{-1}$) are the best fit parameters for E$^{1}_{g}$, E$^{2}_{g}$ and A$_{1g}$ Raman active modes, respectively.}
\label{fig2}
\end{figure}

Importantly, the observed phonon structure and, consequently, the crystal $P\bar{3}c1$ symmetry persist at temperatures $T\gg T_{PT}$ deep in the high temperature regime, as evident from Figure~\ref{fig2}, indicating that these are the characteristics of the high temperature phase.\ As the temperature is increased all the modes are shifted toward lower energies and become progressively broader (see Inset to Fig.~\ref{fig2}).\ All the changes of the spectra induced by temperature increase are in accordance with the well known anharmonicity model.\cite{anh1,Lazarevic1,Lazarevic2}

\begin{figure}
\includegraphics[width=0.45\textwidth]{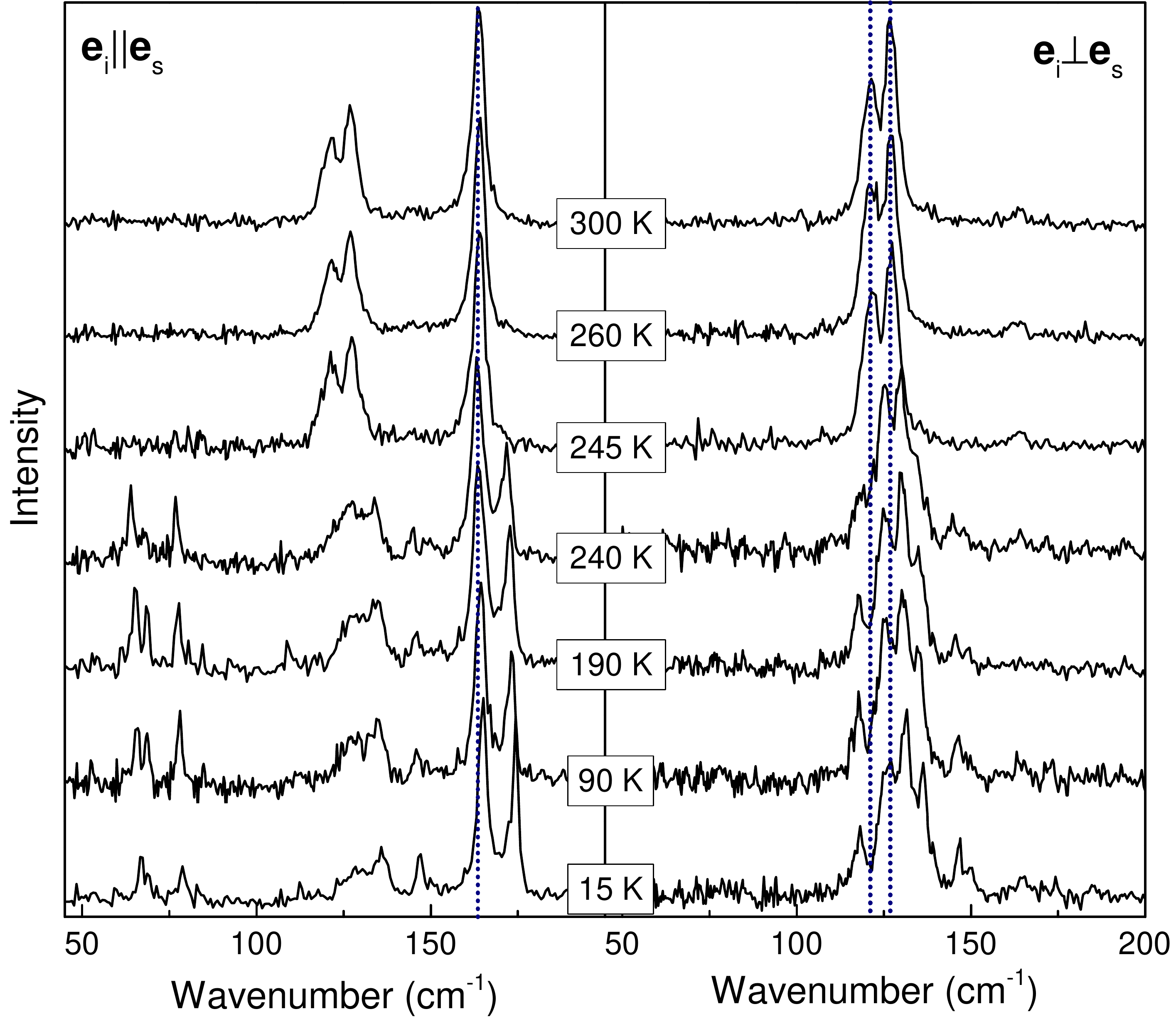}
\caption{(Color online) Polarized Raman scattering spectra of IrTe$_2$ measured at various temperatures, as indicated, in the low temperature regime using JY T64000 Raman system in parallel and cross polarization configurations.\ The spectra were measured by warming the sample from 15 K.\ Dotted vertical lines represent guide to the eyes.}
\label{fig3}
\end{figure}

%
%%%%%%%%%%%%%%%%%%%%%%%%%%%%%%%%%%
\subsection{Low temperature phase}
%%%%%%%%%%%%%%%%%%%%%%%%%%%%%%%%%%
%

By lowering the temperature IrTe$_2$ undergoes the phase transition in the range between 220 K and 280 K.\cite{Trig-Mono,P1,P-1}\ The origin of the phase transition as well as the crystal symmetry of IrTe$_2$ at low temperatures are still under vigorous debate.\cite{Trig-Mono,P1,Zhang,P-1}

Polarized Raman scattering spectra of IrTe$_2$ measured at low temperatures (between 15 K and 300 K) in parallel and cross polarization configurations are presented in Figure~\ref{fig3}.\ Significant changes in the spectra in both polarization configurations are observed around 245 K.\ Lower transition temperature is a consequence of the local heating effects of the sample by the laser beam.\ Unlike in the case of a canonical CDW phase transition where additional modes gradually appear,\cite{LazarevicCDW} observed sudden change in the phonon spectra suggests the first order character of the phase transition.\ Existence of at least eleven peaks in the low temperature phase indicates lowering of the symmetry and/or an increase of the unit cell size.\
\begin{figure}
\includegraphics[width=0.44\textwidth]{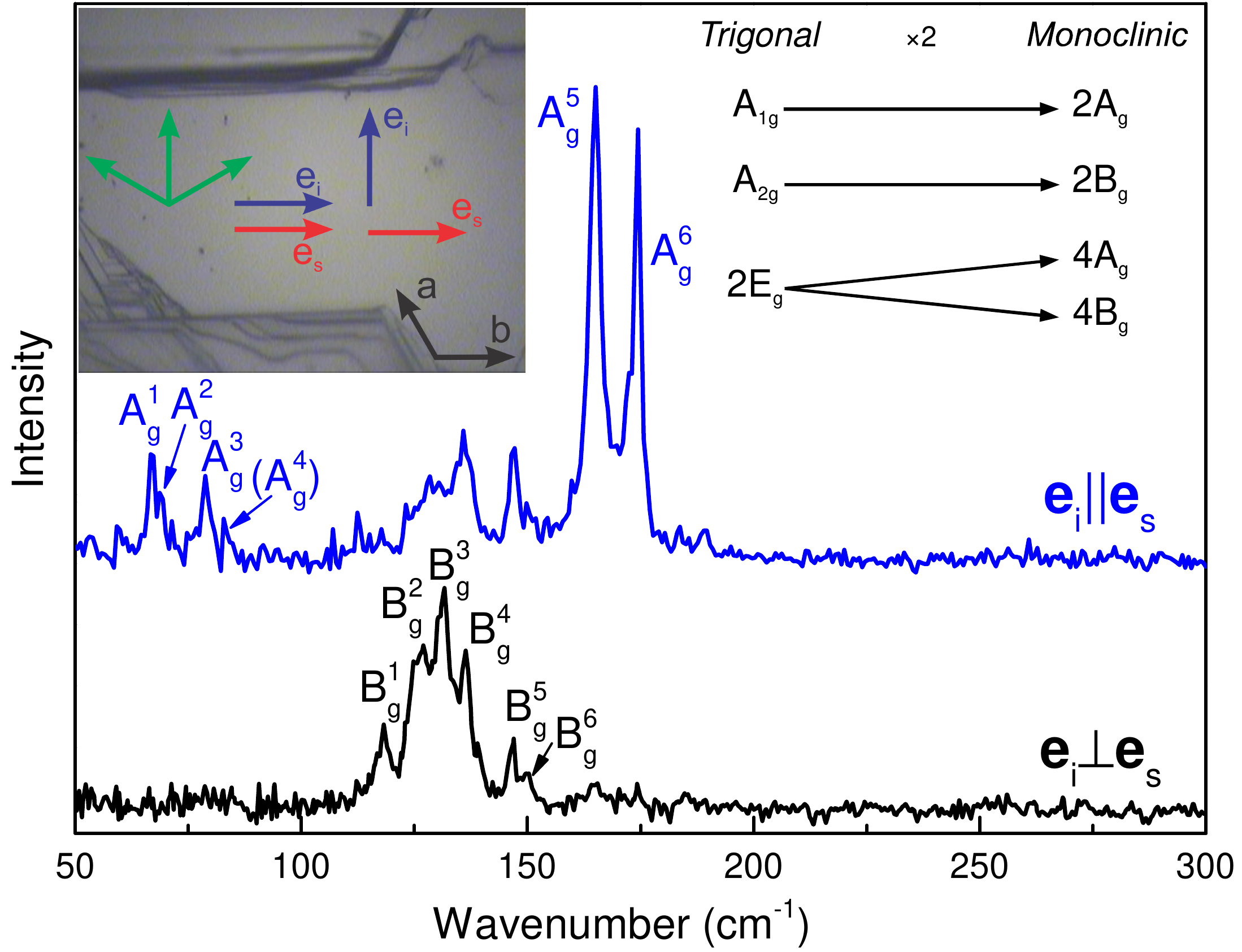}
\caption{(Color online) Polarized Raman scattering spectra of IrTe$_2$ measured at 15 K using JY T64000 Raman system in parallel and cross polarization configurations.\ Inset on the left: Image of the IrTe$_2$ sample.\ Inset on the right: Correlation diagram connecting Raman active phonons for trigonal and monoclinic type of structures.}
\label{fig4}
\end{figure}

\begin{figure}[t!]
\includegraphics[width=0.40\textwidth]{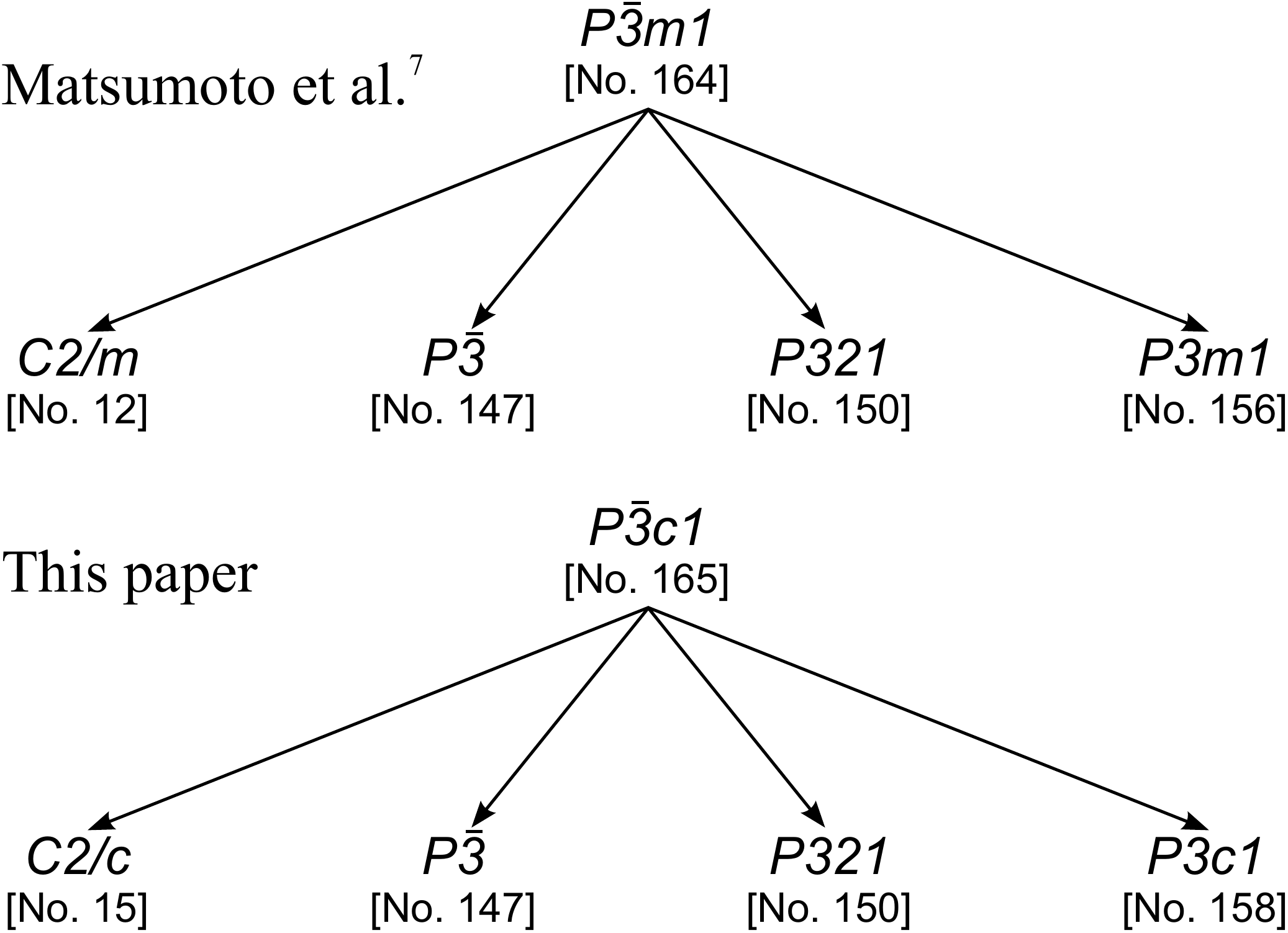}
%\begin{tikzpicture}[node distance=2cm,]
%  \node[align=center] (A)            {$C2/m$ \\ {\scriptsize $[No. 12]$}};
%  \node[align=center] (B)  [right of=A]            {$P\bar{3}$ \\ {\scriptsize$[No. 147]$}};
%  \node[align=center] (C)  [right of=B]            {$P321$ \\ {\scriptsize$[No. 150]$}};
%  \node[align=center] (D)  [right of=C]            {$P3m1$ \\ {\scriptsize$[No. 156]$}};
%  \node[align=center] (E)  [above right=1cm and -0.44cm of B]                  {$P\bar{3}m1$ \\ {\scriptsize$[No. 164]$}};
%  \draw [->] (E.south) -- (A.north);
%  \draw [->] (E.south) -- (B.north);
%  \draw [->] (E.south) -- (C.north);
%  \draw [->] (E.south) -- (D.north);
%  \node[align=center] (E)  [above=1cm of A]                  {Matsumoto et al.\cite{Trig-Mono}};
%\end{tikzpicture}

%\begin{tikzpicture}[node distance=2cm,]
%  \node[align=center] (A)            {$C2/c$ \\ {\scriptsize $[No. 15]$}};
%  \node[align=center] (B)  [right of=A]            {$P\bar{3}$ \\ {\scriptsize$[No. 147]$}};
%  \node[align=center] (C)  [right of=B]            {$P321$ \\ {\scriptsize$[No. 150]$}};
%  \node[align=center] (D)  [right of=C]            {$P3c1$ \\ {\scriptsize$[No. 158]$}};
%  \node[align=center] (E)  [above right=1cm and -0.2cm of B]                  {$P\bar{3}c1$ \\ {\scriptsize$[No. 165]$}};
%  \draw [->] (E.south) -- (A.north);
%  \draw [->] (E.south) -- (B.north);
%  \draw [->] (E.south) -- (C.north);
%  \draw [->] (E.south) -- (D.north);
%  \node[align=center] (E)  [above=1cm of A]                  {This paper};
%\end{tikzpicture}
\caption{Schematics of maximal non-isomorphic subgroup relations of the space group $P\bar{3}m1$\cite{Trig-Mono} (upper panel) and $P\bar{3}c1$ (lower panel) for t-subgroup.\ The number of the space group is given in the parenthesis.}
\label{fig6}
\end{figure}

Figure~\ref{fig4} shows polarized Raman scattering spectra of IrTe$_2$ measured at 15 K in parallel and cross polarization configurations.\ Significant difference of the spectra measured in the parallel and crossed polarization configuration indicates the exitance of the two separate scattering channels in the low temperature phase.\ Symmetry arguments suggest that in the case of triclinic crystal structure only one channel can be observed.\ Due to the proposed orientation of the triclinic lattice\cite{P1,P-1} in relation to the trigonal lattice, the contribution to the scattering intensity (in our scattering geometry) would come from nearly all the components of the Raman tensor (see Table~\ref{tab.1}) and the cancelation of some Raman modes in different polarization configurations is highly unlikely.\ Furthermore, for both $P1$ and $P\bar{1}$ space groups a substantially larger number of Raman modes is expected to be observed in the measured spectra.\ All this suggest that the IrTe$_2$ crystal symmetry in the low temperature phase should be higher than triclinic ($P1$ or $P\bar{1}$).\ The next crystal system with two different scattering channels is monoclinic (see Table~\ref{tab.1}).\ The obtained spectra may be interpreted within the monoclinic crystal symmetry provided that the optical axis of the low temperature phase is orthogonal to direction of incident light in the Raman scattering experiment, \emph{i.e.} if it lies in the (001) plane of the trigonal phase.\ This is consistent with the picture proposed by Matsumoto \emph{et al.}\cite{Trig-Mono} At this point one should have in mind that symmetry breaking may occur along three equivalent directions indicated by the green arrows in the left Inset to Fig.~\ref{fig4}.\ For generality we assume the contributions from all three possible orientations.\ Consequently one may expect the appearance of the B$_g$ modes in both parallel and cross polarization configurations.\ Although the A$_g$ modes may be also observed in both parallel and cross polarization configurations (with the assumption of twinning), in crossed polarization configuration the intensity of the A$_g$ modes depend on the $|b-c|^2$ and the cancelation can be easily achieved.\

Following the previous arguments the peaks at about 67 cm$^{-1}$, 69 cm$^{-1}$, 79 cm$^{-1}$, 165 cm$^{-1}$ and 174 cm$^{-1}$ which can be observed in parallel but not in cross polarization configuration may be assigned as the A$_g$ symmetry modes.\ We believe that weak structure at about 83 cm$^{-1}$ may also be the A$_g$ symmetry mode, however very low intensity prevents unambiguous assignation.\ Six peaks at about 118 cm$^{-1}$, 126 cm$^{-1}$,  131 cm$^{-1}$, 136 cm$^{-1}$, 148 cm$^{-1}$ and 150 cm$^{-1}$ that can be observed in both parallel and crossed polarization configuration are assigned as the B$_g$ symmetry modes.\

\begin{figure}
\includegraphics[width=0.40\textwidth]{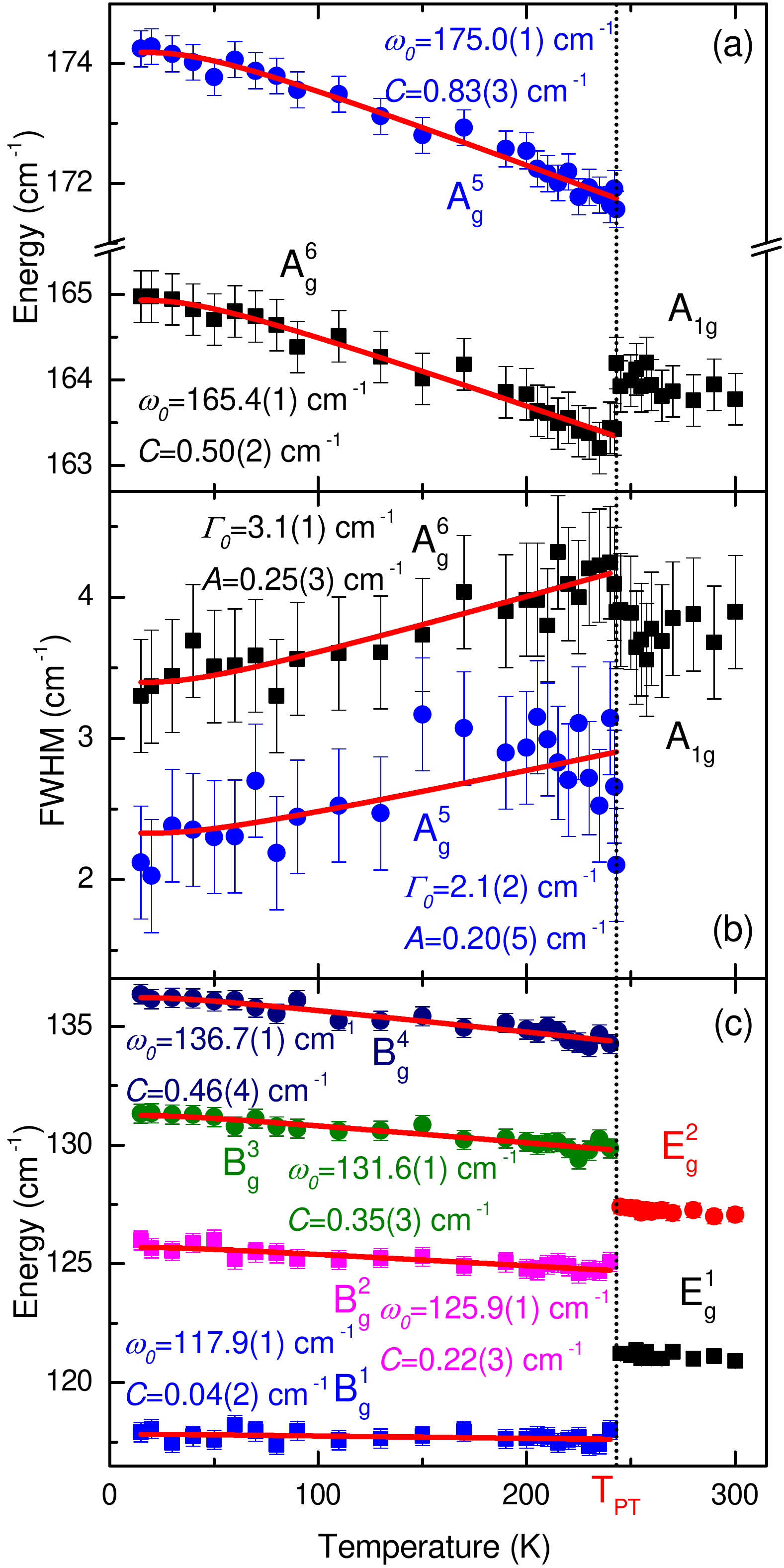}
\caption{(Color online) Temperature dependence of the highest intensity Raman modes energy (a) and (c) and linewidth (b).\ Solid lines represent calculated spectra by using the standard three-phonon anharmonicity model.\cite{anh1} The spectra were measured by warming the sample from the 15 K.}
\label{fig5}
\end{figure}

Although the properties of the observed Raman modes can, in principle, be interpreted within the monoclinic crystal system, proposed\cite{Trig-Mono} unit cell of the $C2/m$ symmetry group with $Z=2$ cannot account for the number of the observed Raman modes.\ According to FGA for $C2/m$ ($Z=2$) only three modes are expected to be observed in the Raman scattering experiment (see Table~\ref{tab.1}).\ Consequently, larger unit cell within the monoclinic crystal system is needed to reproduce the observed Raman spectra.\ Following the previous arguments and the discussion regarding the symmetry of the high temperature phase we may conclude that the space group symmetry of IrTe$_2$ at low temperatures should be searched for within the monoclinic $C2/c$ space group or some of it's t-subgroups (see Figure~\ref{fig6}).\cite{international}

The temperature evolution of the Raman spectra (see Fig.~\ref{fig3}) across the phase transition can be seen as the splitting of the two E$_g$ modes into A$_g$ (A$_g^{1}$--A$_g^{4}$) and B$_g$ (B$_g^{1}$--B$_g^{4}$) quartets due to symmetry lowering and (at least) two times increase of the primitive cell size (see right Inset to Fig.~\ref{fig4}).\ The A$_g^{5}$ and A$_g^{6}$ most likely originate from the A$_{1g}$ mode whereas the B$_g^{5}$ and B$_g^{6}$ originate from the A$_{2g}$ mode of the trigonal phase.\ The absence of the E$_g^1$ and E$_g^2$ modes (within our experimental resolution), characteristic for the trigonal phase, in the low temperature Raman spectra of IrTe$_2$ suggests the absence of the trigonal lattice at low temperatures.\cite{Zhang}

Temperature dependence of the energy and linewidth for the highest intensity Raman modes is shown in Figure~\ref{fig5}.\ Clear fingerprint of the first order phase transition is observed in both the energy and the linewidth of the observed modes.\ Solid lines represent calculated spectra for the low temperature phase by using three-phonon anharmonicity model.\cite{anh1} Good agreement with the experimental data confirms that anharmonicity plays major role in the temperature dependence of the temperature phase phonons self-energy below $T_{PT}$.\

\section{Conclusion}

The Raman scattering study of IrTe$_2$ single crystals has been presented.\ At room temperature, three instead of two Raman active modes predicted by factor group analysis for the $P\bar{3}m1$ symmetry group are observed.\ The Raman data showed that the $P\bar{3}c1$ rather than $P\bar{3}m1$ crystal symmetry is needed to describe the phonon structure of IrTe$_2$ at room temperature.
The sudden change in the Raman spectra below $T_{PT}=$ 245 K revealed the first order structural phase transition.\ The properties of the phonon spectra below $T_{PT}$ are well interpreted within the monoclinic crystal symmetry.\ The splitting of the E$_g$ modes at the $T_{PT}$ comes from the symmetry lowering and the increase in the size of the unit cell.\ We believe that the space group symmetry of IrTe$_2$ at low temperatures should be searched for within the monoclinic $C2/c$ space group or some of it's t-subgroups.\ Further structural investigation of both trigonal and monoclinic phases of IrTe$_2$ are needed.\ Apart from the symmetry change at $T_{PT}$, the temperature dependence of the energy and linewidth of the Raman active modes of IrTe$_2$ are mostly anharmonic.\
\section*{Acknowledgment}
We gratefully acknowledge discussions with R. Hackl. This work was supported by the Serbian Ministry of Education, Science and Technological Development under Projects ON171032 and III45018, as well as Serbian - Germany bilateral project: ''Interplay of Fe-vacancy ordering and spin fluctuations in iron - based high temperature superconductors''. Part of this work was carried out at the Brookhaven National Laboratory which is, as well as the NSLS facility, operated for the Office of Basic Energy Sciences, U.S. Department of Energy by Brookhaven Science Associates, under Contract No. DE-AC02-98CH10886.

%\bibliography{IrTe2-ref}


\end{document}